\documentclass[seceq]{ptptex}

\usepackage{graphicx}
\newcommand{\lk}{\left( }
\newcommand{\rk}{\right)}
\newcommand{\ltk}{\left\{ }
\newcommand{\rtk}{ \right\} }
\newcommand{\ldk}{\left[ }
\newcommand{\rdk}{ \right] }
\title{%        %You can use \\ for explicit line-break
Brane-induced Skyrmions\\
--\hspace{1mm}{\it\small Baryons in Holographic QCD}\hspace{1mm}--
}

\author{%
Kanabu \textsc{Nawa}, 
Hideo \textsc{Suganuma}
and
Toru \textsc{Kojo}
}

\inst{%         %Affiliation, neglected when [addenda] or [errata]
Department of Physics, Kyoto University, 
                Kyoto 606-8502, Japan}

\abst{%         %this abstract is neglected when [addenda] or [errata]
We study baryons 
in holographic QCD with
D4/D8/$\overline{\rm D8}$ multi D brane system.
In holographic QCD, the baryon appears as a topologically non-trivial
chiral soliton in a four-dimensional effective theory of mesons,
which is called `Brane-induced Skyrmion'. 
We derive and calculate the Euler-Lagrange equation for the hedgehog configuration 
with chiral profile $F(r)$ and $\rho$-meson profile $\tilde G(r)$,
and obtain the soliton solution of the holographic QCD.
}
\begin{document}

\maketitle

%\section{Title page}
%Please be aware that there are two fields for the 
%title and authors.  
%One for the {\bf full authors' name}\\
%\hspace*{1cm} Daisuke Jido and Atsushi Hosaka,\\ 
%and the other with initials\\
%\hspace*{1cm} D. Jido and A. Hosaka.\\
%Similarly, one {\bf for full title}\\
%\hspace*{1cm} Recognition of Shapes and Colors at YKIS06 While Thinking\\
%and the other for {\bf short one}\\
%\hspace*{1cm} Recognition of Shapes and Colors

\section{Holographic QCD, color confinement and chiral symmetry breaking}

Based on the recent remarkable progress in the concept
of gauge/gravity duality,
Sakai-Sugimoto succeeded to construct 
%the non-perturbative properties of QCD, especially 
the low-energy theory of massless QCD from 
the multi D-brane system consisting of  
%configurations with 
D4/D8/$\overline{\rm D8}$ 
in type IIA superstring theory~\cite{SS}.
%This is called Sakai-Sugimoto model, which is regarded as one of the reliable
%holographic models.

\begin{figure}[h]
\begin{center}
    \begin{tabular}{cc}
\hspace*{-3mm}
\begin{minipage}{68.5mm}
\begin{center}
       \includegraphics[width=6.0cm]{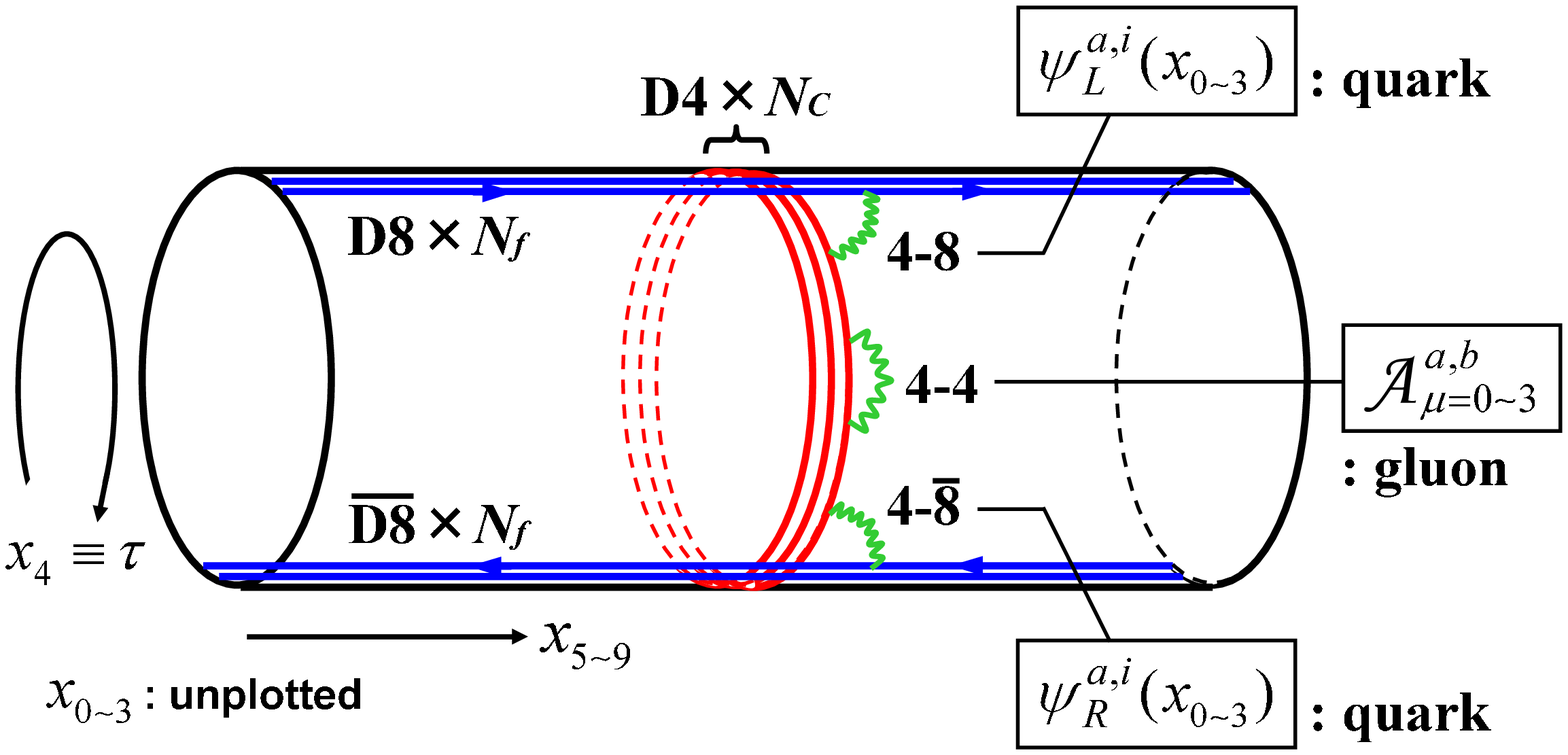}
       \caption{Multi D-brane configurations of the D4/D8/$\overline{\rm D8}$
 holographic QCD. $N_c$-folded D4 branes and $N_f$-folded
 D8-$\overline{\rm D8}$ branes. 
%Flat four-dimensional space-time $x_{0\sim 3}$ are not plotted.
Gluons and quarks appear as 
 4-4, 4-8 and 4-$\overline{8}$ strings shown by waving lines. } 
\label{D4D8D8_1_light}
\end{center}
\end{minipage} &
\hspace{-2mm}
\begin{minipage}{68.5mm}
\vspace{4mm}
\begin{center}
       \includegraphics[width=6.0cm]{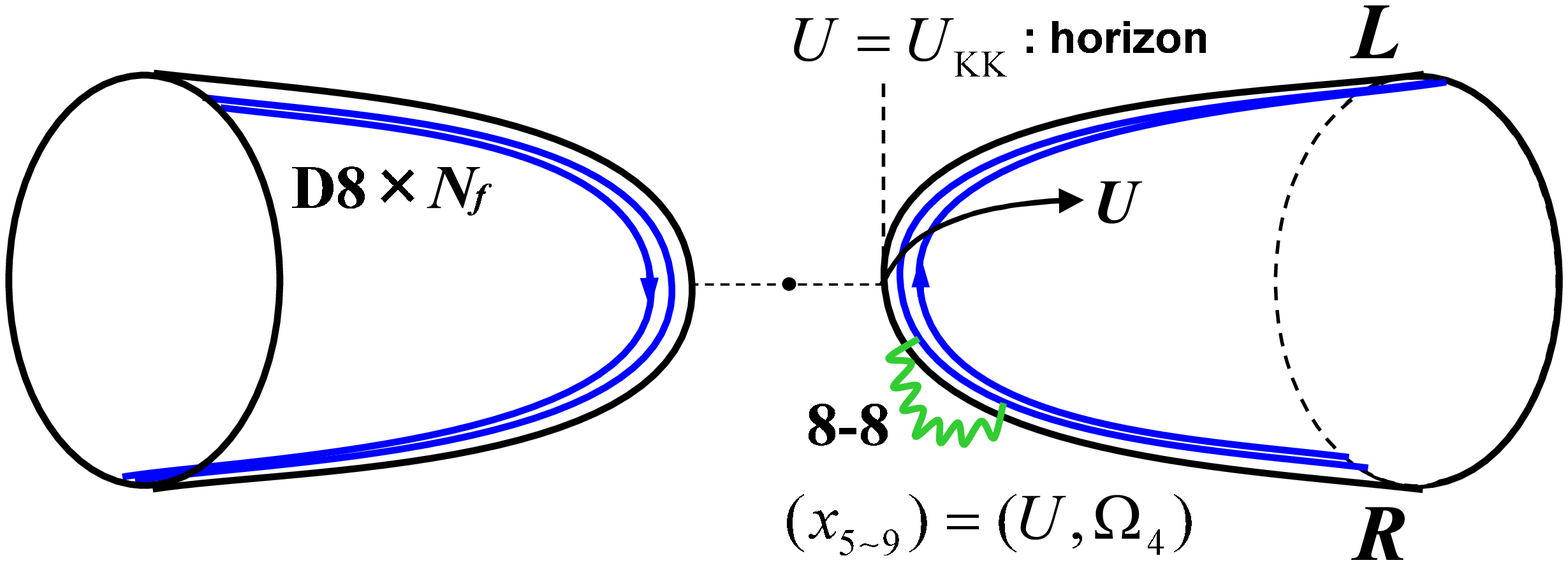}
       \caption{Probe D8 branes with D4 supergravity background.
%Flat four-dimensional space-time $x_{0\sim 3}$ are not plotted.
The radial coordinate $U$ in the extra five dimensions $x_{5\sim 9}$
is bounded from below by a horizon $U_{\rm KK}$ as $U\geq U_{\rm KK}$.
Color-singlet mesons appear from 8-8 strings shown by the waving line.}
\label{D8_backD4_1_light}
\end{center}
\end{minipage}
    \end{tabular}\par
  \end{center}
\end{figure}

In the D4/D8/$\overline{\rm D8}$ holographic QCD, 
D4/D8/$\overline{\rm D8}$-branes are placed as Table \ref{D4D8_conf}, 
and D4 branes are compactified along an extra direction $x_4\equiv\tau$ 
with the Kaluza-Klein mass scale as 
$\tau \sim \tau+2\pi M_{\rm KK}^{-1}$,
to take away the extra-modes for massless QCD.
The compositions of the four-dimensional massless QCD,
{\it i.e.}, 
gluons and massless quarks, are represented as the fluctuation modes of 
open strings on 
$N_c$-folded D4 branes and 
$N_f$-folded D8 and $\overline{\rm D8}$ branes as 
Fig.\ref{D4D8D8_1_light}.
In the D4/D8/$\overline{\rm D8}$ holographic QCD with large $N_c$, 
D4 branes are represented by the classical supergravity background 
with the concept of `gauge/gravity duality', 
and back reaction from the probe D8 and $\overline{\rm D8}$ branes
to the total system can be neglected as a probe approximation.
%
%--------------
%--------------
%--------------
\vspace{-4mm}
\begin{table}[h]
%\hspace*{13mm}
\begin{minipage}{140mm}
\begin{center}
\begin{tabular}{ccccccccccc}
 {} & 0 & 1 & 2 & 3 & 4 & 5 & 6 & 7 & 8 & 9 \\ \hline
 D4 & $\bigcirc$ & $\bigcirc$ & $\bigcirc$ & $\bigcirc$ & $\bigcirc$ & %
    {} & {} & {} & {} & {}\\
 D8$\mbox{-}\overline{\rm D8}$ & $\bigcirc$ & $\bigcirc$ & $\bigcirc$ & $\bigcirc$ & %
    {} & $\bigcirc$ & $\bigcirc$ & $\bigcirc$ & $\bigcirc$ & $\bigcirc$\\ 
\end{tabular}
\caption{The space-time extension of the D4 brane and
 D8-$\overline{\rm D8}$ branes to construct massless QCD.  
The circle denotes the extended direction of each D brane.
$x_{0 \sim 3}$ correspond to flat space-time.}
\label{D4D8_conf}
\end{center}
\end{minipage}
\end{table}
%---------------
%---------------
%---------------

\vspace{-2mm}
%Since the mass of D brane is proportional to the folding number of the D brane
%as the Ramond-Ramond flux quantum,
%%therefore there occurs the idea of describing the
%the existence of the D brane with large folding number
%can be described by a curved space-time.
%
%Thus the supergravity description of D brane and also the 
%construction of gauge theory from the D brane eventually gives the concept of `duality'
%between the supergravity and gauge theory as the gauge/gravity duality
%mediated by the mutual D brane.

In the presence of $N_c$-folded D4 brane, 
the radial direction $U$ in extra-dimensions $x_{5\sim 9}$
is bounded from below like $U\geq U_{\rm KK}$ as a `horizon' in
%there appears a horizon $U_{\rm KK}$ in
ten-dimensional space-time. 
%as a classical D4 supergravity solution.
%
Then,
probe D8 and $\overline{\rm D8}$ branes are interpolated with each
other in this curved space-time as in Fig.~\ref{D8_backD4_1_light}.
This interpolation indicates 
symmetry breaking of 
${\rm U}(N_f)_{\rm D8}\times {\rm U}(N_f)_{\overline{\rm D8}}$
%symmetry
into 
%the single-valued 
${\rm U}(N_f)_{\rm D8}$, 
%symmetry,
which can be regarded as the holographic manifestation of spontaneous chiral symmetry
breaking.

Furthermore, color confinement is realized in the holographic QCD as follows.  
Since color quantum number is carried only by $N_c$-folded D4 branes,
colored objects appear as the fluctuation modes of
open strings with at least one end located
on $N_c$-folded D4 branes, {\it e.g.}, 
gluons from 4-4 strings and 
%flavor 
quarks from 4-8 strings.
Therefore, in the supergravity background of D4 brane,
%it can be interpreted that 
these colored objects
appearing around D4 branes
would locate behind the horizon $U_{\rm KK}$
and become invisible from outside,
which can be interpreted as color confinement at 
%the
low-energy scale.
%figure---------------------------------------------
\vspace{-3.5mm}
%-----------------------------------------

\vspace{-1.5mm}
From these considerations, one may see that
%These considerations seems to suggest that
chiral symmetry breaking and color confinement occur simultaneously;
two independent chirality spaces are connected by the `worm hole' into
which colored objects are absorbed as shown in Fig.\ref{D8_backD4_1_light}.

After the supergravity description of background D4 branes,
there only appear colorless objects as the fluctuation modes of
residual probe D8 branes:
mesons and also baryons.
Since large-$N_c$ QCD becomes equivalent with the 
weak-coupling system of mesons and glueballs~\cite{tHooft},
baryons do not directly appear as the dynamical degrees of freedom
%
%Therefore, in this large-$N_c$ holographic QCD,
%baryons are expected to 
but appear 
%not directly but
as some soliton-like topological objects~\cite{Skyrme} 
in this large-$N_c$ holographic QCD.
%
%In our investigation,
In this paper, 
we study the baryon as a topologically non-trivial chiral soliton, 
which is called as `Brane-induced Skyrmion', 
in the four-dimensional meson effective theory
induced by the D4/D8/$\overline{\rm D8}$ holographic QCD.~\cite{NSK}

The existence of D4 supergravity background is reflected on the metric 
of the non-abelian Dirac-Born-Infeld (DBI) action of D8 brane
in nine-dimensional space-time.
After integrating out 
%the extra four-dimensional coordinate space,
%the symmetric directions along the
extra four-dimensional angular coordinates around $x_{5 \sim 9}$,
the effective action of D8 brane is reduced into the 
five-dimensional Yang-Mills theory with flat four-dimensional space-time
$(t,{\bf x})$ and other extra fifth dimension $z$ with curved measure
as~\cite{SS}
\begin{eqnarray}
S_{\rm D8}^{\rm DBI}=
%-S_{\rm D8}^{\rm DBI}|_{A_M\rightarrow 0}&=&
\kappa\int d^4 x dz\mbox{tr}\ltk
\frac{1}{2}K(z)^{-\frac{1}{3}}F_{\mu\nu}F_{\mu\nu}+K(z) F_{\mu z}F_{\mu z}\rtk+O(F^4),
\label{5dimDBI_2}
\end{eqnarray}
where $A_M$-independent part is abbreviated.
%
%Here, 
$K(z)\equiv 1+z^2$ is the non-trivial curvature
in the fifth direction $z$ induced by the supergravity background of
D4 branes, and 
$\kappa\equiv \frac{\lambda N_c}{108\pi^3}$ 
with the 't~Hooft coupling $\lambda\equiv g_{\rm YM}^2N_c$.
%For simplicity, 
We here take $M_{\rm KK}=1$ unit.
We now treat 
non-trivial leading order $O(F^2)$ of the DBI action above,
corresponding to the leading order of $1/N_c$ and $1/\lambda$ expansions 
in the holographic model, 
%with respect to the 't~Hooft coupling expansion
%along the discussions below
for the argument of non-perturbative (strong coupling) properties of QCD.
%This leading order is derived and uniquely determined in the holographic model.

%For the argument of non-perturbative (strong coupling) properties of QCD, 
%we treat non-trivial leading order $O(F^2)$ of the DBI action,
%which is the uniquely-determined leading order of $1/N_c$ and $1/\lambda$ expansions
%in the holographic QCD.

To obtain the four-dimensional effective theory
with definite parity and G-parity of QCD,
we perform the mode expansion of the five-dimensional gauge field $A_M(x_N)$
($M,N=0 \sim 4$)
with respect to the extra-coordinate $z$ 
%the four-dimensional parity eigen-states 
by using proper 
parify-definite orthogonal basis $\psi_\pm(z)$ and $\psi_{n}(z)$
$(n=1,2,\cdots)$ as 
%in the $z$ direction 
%
\begin{eqnarray}
& &A_\mu(x_N) 
=l_\mu(x_\nu)\psi_+(z)+
   r_\mu(x_\nu)\psi_-(z)+
 \sum_{n\geq 1}B_\mu^{(n)}(x_\nu)\psi_n(z),\label{limit2_lr} \\
%\end{eqnarray}
%\vspace{-9mm}
%\begin{eqnarray}
& & l_\mu(x_\nu) \equiv \frac{1}{i}\xi^{-1}(x_\nu)\partial_\mu\xi (x_\nu),\hspace{2mm}
     r_\mu(x_\nu)\equiv \frac{1}{i}\xi
     (x_\nu)\partial_\mu\xi^{-1}(x_\nu),\hspace{2mm}
 \xi(x_\nu)\equiv e^{i\pi(x_\nu)/f_\pi},~~~~~~
\label{lr_def}
\end{eqnarray}
%
%supporting all parts of the gauge field $A_\mu$ at
%$z\rightarrow \pm\infty$:
%$\psi_\pm(z\rightarrow \pm\infty)=1$ and
%$\psi_\pm(z\rightarrow \mp\infty)=0$.
%
where `$A_z=0$ and $\xi_+^{-1}=\xi_-$' gauge is taken.~\cite{SS,NSK} 
To diagonalize the five-dimensional Yang-Mills theory with curved
measure $K^{-\frac{1}{3}}(z)$ and $K(z)$ in the fifth direction $z$,
the basis $\psi_{n}(z)$ are taken to be 
the normalizable eigen-function satisfying
%
%\vspace{-1mm}
\begin{eqnarray}
-K(z)^{\frac{1}{3}}\frac{d}{dz}\ltk K(z) \frac{d\psi_n}{dz}\rtk =
 \lambda_n\psi_n,
\hspace{7mm}\mbox{($ 0< \lambda_1 < \lambda_2 < \cdots$)}\label{eigen1}
\end{eqnarray}
and the basis $\psi_\pm(z)\equiv \frac{1}{2}\pm \frac{1}{\pi}\arctan z$
are taken as zero modes of this eigen-equation.
%
%\vspace{-2mm}
%\hspace{-7mm}
%with normalization condition:
%$\kappa\int dz K(z)^{-\frac{1}{3}}\psi_m \psi_n =\delta_{nm}$.

In the holographic QCD,
$B_\mu^{(n=1,2,\cdots)}$ are regarded as 
infinite number of excitation modes
of
(axial-)vector mesons. 
%like $\rho_\mu\equiv B_\mu^{(1)}, a_{1,\mu}\equiv B_\mu^{(2)},
%\rho^{'}_\mu\equiv B_\mu^{(3)}, {\it etc}$.
%
Here, the basis $\psi_\pm$ and $\psi_{n}$ can be regarded
as the `wave-function' of pions and (axial-)vector mesons
in the fifth dimension.
The mass of (axial-)vector mesons is given
by the eigen-value of the basis $\psi_{n}$ in Eq.(\ref{eigen1})
as $m_n=\lambda_n$ in this holographic QCD,
which indicates that
a large oscillation in the extra fifth direction 
induces a large mass in the four-dimensional theory.
As for pions,
slightly oscillating component $\psi_\pm$ of pion wave functions
are, in fact, the zero mode of Eq.(\ref{eigen1}) with curved measure,
so that pions appear as massless objects, corresponding
to the `geodesic' of curved five-dimensional space-time.~\cite{NSK}

The smaller overlapping of wave functions in the extra fifth direction 
between pions and `largely oscillating' heavier (axial-)vector mesons 
predicts 
the smaller coupling constant between them~\cite{NSK,DTSon} 
through the projection of the action 
into flat four-dimensional space-time.
%which is numerically checked by $z$ integral.
%
Therefore, for the study of chiral solitons, 
which consist of large-amplitude pion fields, 
we can expect smaller effects from 
%higher mass excitation modes of 
heavier (axial-)vector mesons, 
so that it would be enough to consider only pions and 
$\rho$ mesons as 
the lowest massive mode $B_\mu^{(1)}$ ($\equiv \rho_\mu$)
in the holographic QCD~\cite{NSK}.

\section{Brane-induced Skyrmion and its properties}

Now, substituting the mode expansion of the gauge
field (\ref{limit2_lr})
in the five-dimensional Yang-Mills theory,
we get the four-dimensional effective Lagrangian
with pions and $\rho$ mesons from holographic QCD,
without small amplitude expansion, as
%===========================================
\begin{eqnarray}
& &{\cal L}_{\rm D8}^{\rm DBI} =\kappa\int dz \mbox{tr}\ltk
               \frac{1}{2}K(z)^{-\frac{1}{3}}F_{\mu\nu}F_{\mu\nu}+
               K(z) F_{\mu z}F_{\mu z}\rtk \nonumber\\
            &=&\frac{f_{\pi}^2}{4}
               \mbox{tr}\lk L_{\mu}L_{\mu}\rk
             - \frac{1}{32 e^2} 
               \mbox{tr}\ldk L_{\mu}, L_{\nu}\rdk^2
%                \nonumber\\ 
%
             + \frac{1}{2}%
               \mbox{tr}\lk
               \partial_{\mu}\rho_{\nu}-\partial_{\nu}\rho_{\mu}\rk^2
             + m_{\rho}^2
               \mbox{tr}\lk\rho_{\mu}\rho_{\mu}\rk
               \nonumber\\
            &+i& g_{3\rho} 
               \mbox{tr}\ltk
               \lk\partial_{\mu}\rho_{\nu}-\partial_{\nu}\rho_{\mu}\rk
               \ldk \rho_{\mu}, \rho_{\nu}\rdk
               \rtk
             - \frac{1}{2}g_{4\rho}
               \mbox{tr}\ldk
               \rho_{\mu}, \rho_{\nu}\rdk^2
%               \nonumber\\
%
            -i g_1
               \mbox{tr}\ltk
               \ldk \alpha_{\mu}, \alpha_{\nu}\rdk
               \lk\partial_{\mu}\rho_{\nu}-\partial_{\nu}\rho_{\mu}\rk
               \rtk
               \nonumber\\
           &+& g_2
               \mbox{tr}\ltk
               \ldk \alpha_{\mu}, \alpha_{\nu}\rdk
               \ldk \rho_{\mu}, \rho_{\nu}\rdk
               \rtk
%              \nonumber\\
%
             + 2 g_3
               \mbox{tr}\ltk
               \ldk \alpha_{\mu}, \alpha_{\nu}\rdk
%               \lk
               \ldk \beta_{\mu}, \rho_{\nu}\rdk 
%               +\ldk \rho_{\mu}, \beta_{\nu}\rdk
%               \rk
               \rtk
%              \nonumber\\
%
            +i 2 g_4
               \mbox{tr}\ltk
               \lk\partial_{\mu}\rho_{\nu}-\partial_{\nu}\rho_{\mu}\rk
%               \lk
               \ldk \beta_{\mu}, \rho_{\nu}\rdk 
%               +\ldk \rho_{\mu}, \beta_{\nu}\rdk
%               \rk
               \rtk
               \nonumber\\
           &-2& g_5
               \mbox{tr}\ltk
               \ldk \rho_{\mu}, \rho_{\nu}\rdk
%               \lk
               \ldk \beta_{\mu}, \rho_{\nu}\rdk 
%               +\ldk \rho_{\mu}, \beta_{\nu}\rdk
%               \rk
               \rtk
%              \nonumber\\
%
             - \frac{g_6}{2} 
               \mbox{tr}\lk
               \ldk \alpha_{\mu}, \rho_{\nu}\rdk +
               \ldk \rho_{\mu}, \alpha_{\nu}\rdk
               \rk^2
%               \nonumber\\
%
             - \frac{g_7}{2} 
               \mbox{tr}\lk
               \ldk \beta_{\mu}, \rho_{\nu}\rdk +
               \ldk \rho_{\mu}, \beta_{\nu}\rdk
               \rk^2~~~~~~
               \label{f11_1}
\end{eqnarray}
with $\alpha_\mu(x_\nu)\equiv l_\mu(x_\nu)-r_\mu(x_\nu)$,
$\beta_\mu(x_\nu) \equiv \frac{1}{2}\ltk l_\mu(x_\nu)+r_\mu(x_\nu)\rtk$ and
$L_\mu = \xi^{-1}\alpha_\mu \xi$.
As a remarkable fact, 
constants $f_\pi$, $m_\rho$, $e$, $g_{3\rho}$, $g_{4\rho}$ and $g_{1\sim 7}$
can be written by the basis $\psi_\pm$ and $\psi_1$, {\it e.g.}, 
$g_{3\rho}\equiv \kappa\int dz K(z)^{-\frac{1}{3}}\psi_1^3$.
The holographic QCD has, in fact, just two parameters $\kappa$ and $M_{\rm KK}$,
%
%so that,
and therefore 
%then 
all the coupling constants in the action
(\ref{f11_1}) are uniquely determined, 
by fixing two parameters 
like experimental inputs for $f_\pi$ and $m_\rho$.
Note also that, because of $g_{3\rho}^2 \ne g_{4\rho}$, the $\rho$-meson part 
in the four-dimensional effective theory differs from the massive Yang-Mills theory.

For the static soliton solution of the action (\ref{f11_1}), 
we take the hedgehog configuration Ansatz
for pion field $U({\bf x})$ and $\rho$ meson field $\rho_\mu ({\bf x})$, 
\begin{eqnarray}
U^{\star}({\bf x})=e^{i\tau_a \hat{x}_a F(r)}, 
\hspace{5mm}
%\mbox{($\hat{x}_a\equiv \frac{x_a}{r}$, $r\equiv|{\bf x}|$,
%$\tau_a$: Pauli matrix)}\label{HH}\\
\rho^{\star}_{0}({\bf x})=0, \hspace{5mm} \rho^{\star}_{i}({\bf x})=\rho^{\star}_{i a}({\bf x}) \frac{\tau_a}{2}
                                               =\ltk \varepsilon_{i a
                                               b}\hat{x}_b\tilde{G}(r)\rtk
                                               \tau_a,
%                                               \hspace{5mm}(\tilde{G}(r)\equiv
%                                               G(r)/r)
\label{WYTP}
\end{eqnarray}
where $F(r)$ ($r \equiv |{\bf x}|$) is a dimensionless function 
with boundary conditions $F(0)=\pi$ and $F(\infty)=0$, giving
topological charge equal to unity as a unit baryon number.

We derive the energy density $\varepsilon [F(r), \tilde{G}(r)]$ of the Brane-induced Skyrmion as 
\begin{eqnarray}
%E[F(r), \tilde{G}(r)]\equiv 
%             \ldk S_{\rm D8}^{\rm DBI}-S_{\rm D8}^{\rm DBI}|_{A_M\rightarrow 0}
%             \rdk_{{\rm hedgehog}}
%                     \equiv \int_0^{\infty}4\pi dr r^2\varepsilon [F(r),
%                     \tilde{G}(r)], \label{BIS_energy}\\
%
& &r^2\varepsilon [F(r), \tilde{G}(r)] =
             \frac{f_\pi^2}{4}\ldk 2\lk r^2 F^{'2}+2\sin^2F \rk\rdk
          + \frac{1}{32e^2}\ldk 16 \sin^2F \lk 2 F^{'2}+\frac{\sin^2F}{r^2}\rk\rdk 
\nonumber\\
         &+& \frac{1}{2}%
             \ldk 8\ltk
             3\tilde{G}^2+2 r \tilde{G}(\tilde{G}^{'}) +r^2\tilde{G}^{'2}\rtk\rdk
          + m_{\rho}^2%
             \ldk 4 r^2 \tilde{G}^2\rdk
          - g_{3\rho}  %
             \ldk 16 r \tilde{G}^3 \rdk
          + \frac{1}{2}g_{4\rho}%
             \ldk 16 r^2 \tilde{G}^4 \rdk
\nonumber\\
         &+&  g_1 %
             \ldk 16 \ltk F^{'}\sin F \cdot\lk
             \tilde{G}+ r \tilde{G}^{'}\rk%
             +\sin^2F\cdot\tilde{G}/r\rtk\rdk
          - g_2%
             \ldk 16 \sin^2F\cdot\tilde{G}^2 \rdk
\nonumber\\
         &-& g_3 %
             \ldk 16 \sin^2F\cdot\lk 1-\cos F\rk \tilde{G}/r \rdk 
          - g_4%
             \ldk 16 \lk 1-\cos F\rk\tilde{G}^2\rdk   
          + g_5 %
             \ldk 16 r \lk 1-\cos F \rk \tilde{G}^3\rdk
\nonumber\\
          &+& g_6%
             \ldk 16 r^2 F^{'2}\tilde{G}^2\rdk
          + g_7%
             \ldk 8 \lk 1-\cos F\rk^2\tilde{G}^2\rdk.
\label{energy_dense_An}
\end{eqnarray}  

%Here we comment on the scaling property of the 
We note that Brane-induced Skyrmion has a scaling property.~\cite{NSK} 
By rewriting energy and length in 
`Adkins-Nappi-Witten (ANW) unit'~~\cite{ANW}
as $E_{\rm ANW}\equiv\frac{f_\pi}{2e} (\propto\kappa M_{\rm KK})$
and $r_{\rm ANW}\equiv\frac{1}{e f_\pi} (\propto \frac{1}{M_{\rm KK}})$, 
%we find that 
all the 
%effects of 
physical parameters like
$f_\pi, m_\rho, e$ 
(and also $m_{a_1}, m_{\rho'},
m_{a_1'},\cdots$
%, if these mesons are included,
)
are uniquely determined 
%extracted 
by the unit of $E_{\rm ANW}$ and $r_{\rm ANW}$,
because the holographic QCD has just two parameters, $\kappa$ and $M_{\rm KK}$.
This scaling property of Brane-induced Skyrmion 
is a remarkable consequence of holographic framework.

\begin{figure}[h]
\begin{center}
    \begin{tabular}{cc}
\hspace*{-25mm}
\begin{minipage}{30mm}
\begin{center}
       \includegraphics[width=5.5cm]{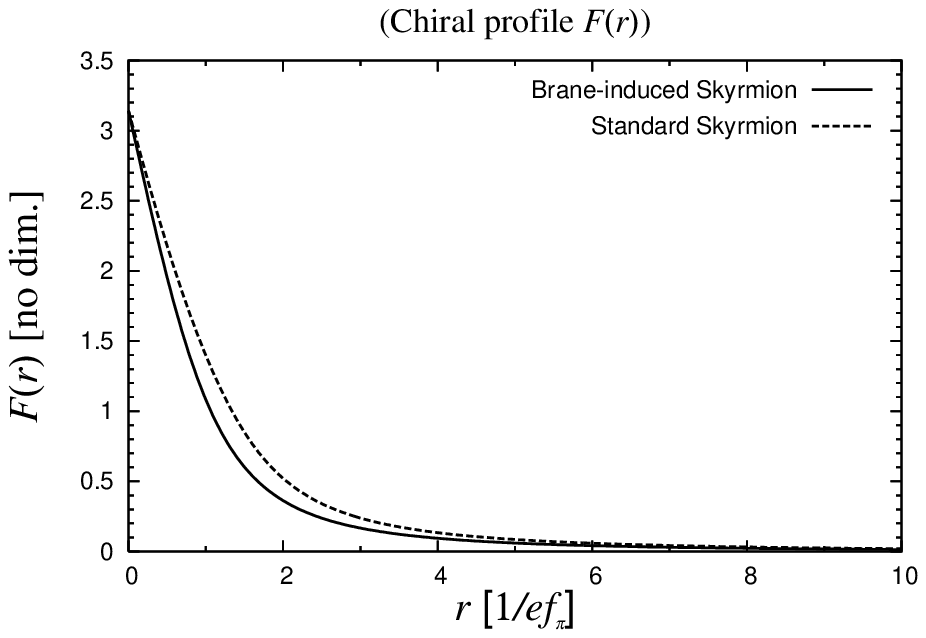}
\end{center}
\end{minipage} &
\hspace{32mm}
\begin{minipage}{30mm}
\begin{center}
       \includegraphics[width=5.5cm]{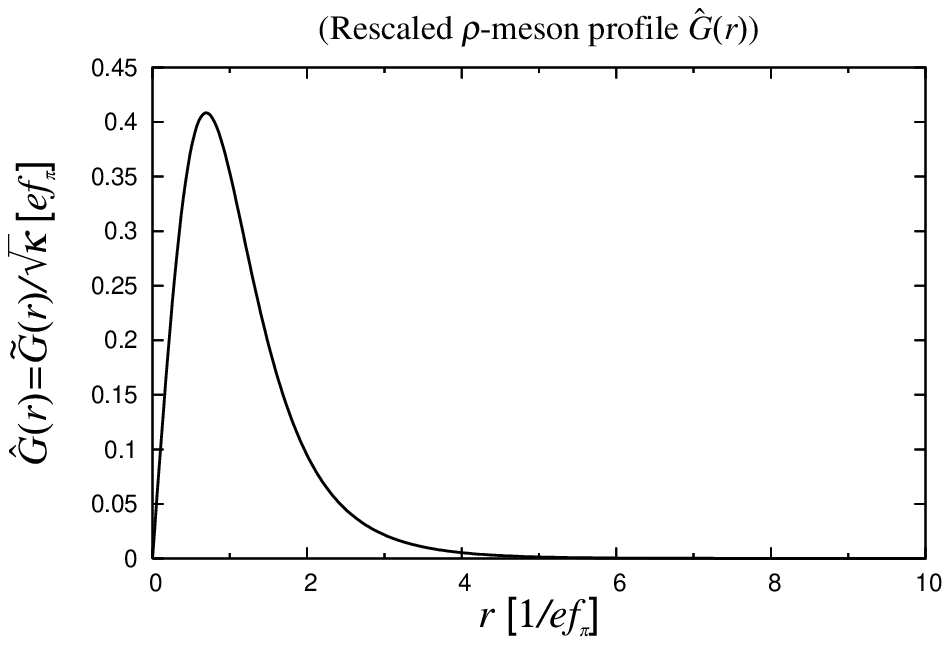}
\end{center}
\end{minipage}
    \end{tabular}\par
\caption{The chiral profile $F(r)$ and rescaled $\rho$-meson profile
 $\widehat{G}(r)$ of Brane-induced Skyrmion as the hedgehog soliton
 solution in the holographic QCD.
 The dashed curve in the left figure denotes the chiral profile of
 standard Skyrmion without $\rho$ mesons.}\label{fig_conf1}
  \end{center}
\vspace{-1mm}
\end{figure}

%By numerically solving the action (\ref{f11_1}) 
With the hedgehog configuration,
we study baryons as 
Brane-induced Skyrmions 
in the large-$N_c$ holographic QCD, 
and obtain the stable soliton solution 
with chiral profile $F(r)$ and rescaled $\rho$-meson profile
$\widehat{G}(r) \equiv\frac{1}{\sqrt{\kappa}}\tilde{G}(r)$,
%rescaling allows scale-invariant analysis for $\rho$ meson 
%configuration~\cite{NSK}
as shown in Fig.\ref{fig_conf1}.
%
%Through the interactions between pions and $\rho$ mesons in the
%holographic QCD, 
%pion contribution seems to be  
%slightly replaced by 
%$\rho$ meson degrees of freedom, 
%which results in the
%shrinkage of $F(r)$ from the 
%standard Skyrme configuration as in Fig \ref{fig_conf1}.
%
%figure---------------------------------------------
\vspace{-4mm}
%----------------------------------------------------------
%
We numerically obtain the total energy (mass) of Brane-induced Skyrmion in ANW unit as
$E\simeq 1.115\times 12\pi^2\ldk \frac{f_{\pi}}{2e}\rdk$
(c.f. 
%which is compared with that of the standard Skyrmion without $\rho$ mesons:
$E\simeq 1.231\times 12\pi^2 \ldk\frac{f_\pi}{2e}\rdk$
for standard Skyrmion~\cite{ANW}).
%given by Adkins {\it et al.}~\cite{ANW}
%
%The result suggests that 
The mass of 
Brane-induced Skyrmion is reduced by $\sim 10\%$ relative to the
standard Skyrmion because of the interactions between pions and $\rho$
mesons
in the meson effective action (\ref{f11_1}) 
or the energy density (\ref{energy_dense_An})
induced by the holographic QCD.

\begin{figure}[h]
\begin{center}
    \begin{tabular}{cc}
\hspace*{-2mm}
\begin{minipage}{67mm}
\begin{center}
       \includegraphics[width=5.5cm]{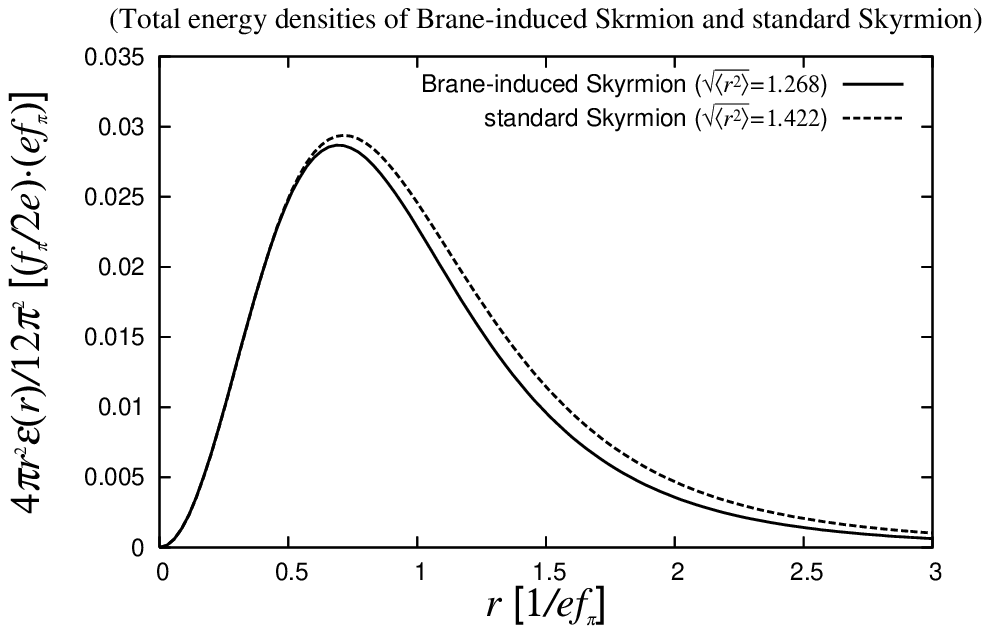}
       \caption{
The energy density profile $4\pi r^2 \varepsilon(r)$ 
(per BPS value of $12\pi^2$) 
of Brane-induced Skyrmion with 
that of standard Skyrmion.}
%Their root-mean-square radii in ANW unit are also presented.}
\label{fig_totalEdense1}
\end{center}
\end{minipage} &
\hspace{-1mm}
\begin{minipage}{68mm}
\begin{center}
       \includegraphics[width=5.5cm]{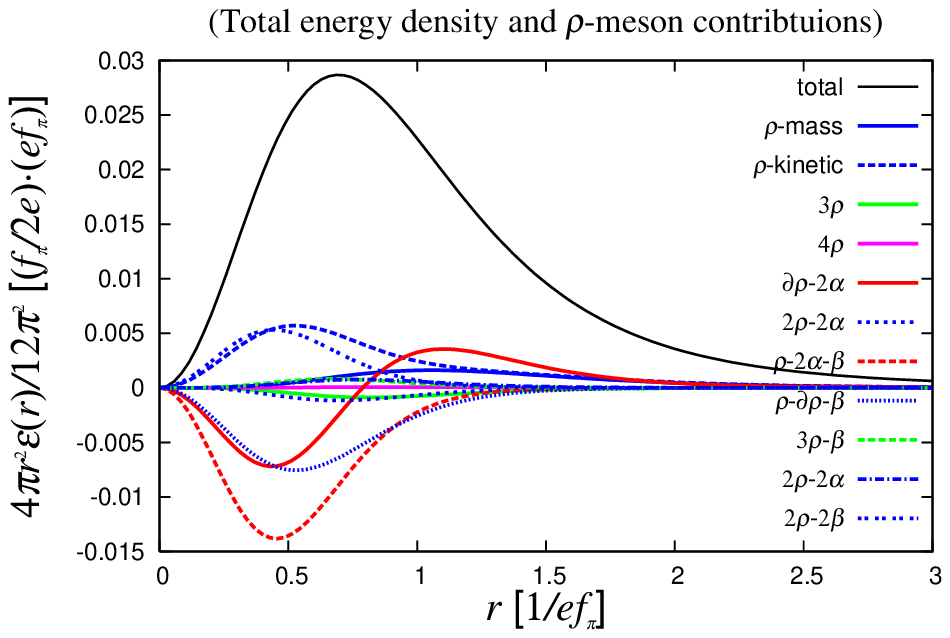}
       \caption{
Contributions of 
$\rho$-meson interaction terms 
in 
%the effective action 
(\ref{f11_1}) 
to the energy density 
%$4\pi r^2 \varepsilon(r)/12\pi^2$ 
%(\ref{energy_dense_An}) 
of Brane-induced Skyrmion
with total energy density.
%Several colors of lines classify the power of $\rho$-meson fields in each
%term.
}
\label{fig_eachEdense2}
\end{center}
\end{minipage}
    \end{tabular}\par
  \end{center}
\end{figure}

Figure~\ref{fig_totalEdense1} shows 
the energy density of Brane-induced Skyrmion. 
%and standard Skyrmion without $\rho$ mesons.
%
%We calculate the root-mean-square radius of the Skyrme configuration
%by using the normalized energy density 
%$\overline{\varepsilon}(r)\equiv \varepsilon(r)/E$
%($\varepsilon(r)$ is total energy density and $E$ is total energy of a
%Skyrmion) as
%$\sqrt{\langle r^2\rangle}\equiv\ldk\int 4\pi r^2 dr
%\overline{\varepsilon}(r)r^2 \rdk^{\frac{1}{2}}$.
%
From the energy-density distribution,
we estimate the 
%root-mean-square `mass' 
radius of the Brane-induced Skyrmion in ANW unit as
$\sqrt{\langle r^2\rangle}\simeq 1.268 \ldk \frac{1}{ef_\pi}\rdk$
%
%which is compared with that of the standard Skyrmion: 
(c.f. $\sqrt{\langle r^2\rangle}\simeq 1.422 \ldk \frac{1}{ef_\pi}\rdk$ for standard Skyrmion).
In the Brane-induced Skyrmion, 
some part of total mass is carried by the heavy $\rho$-meson 
in the soliton core,
which gives the shrinkage of the total size by $\sim 10 \%$
relative to the standard Skyrmion. 
%as in Fig.\ref{fig_totalEdense1}.
%
%The comparison between total energy density and 
We show $\rho$-meson contributions to 
the energy density (\ref{energy_dense_An})
in the Brane-induced Skyrmion in
Fig.\ref{fig_eachEdense2},
which indicates that
$\rho$-meson components are rather active in the core region of
baryons through various interaction terms in the action (\ref{f11_1}).
This active $\rho$-meson component inside baryons may be 
a new striking picture for baryons suggested from the holographic QCD.
%figure---------------------------------------------
\vspace{-1mm}
%-------------------------------------

\vspace{-1mm}
%Finally we recover the physical unit for the mass and
%root-mean-square radius of Brane-induced Skyrmion.
%
By taking the experimental inputs 
%for $f_\pi$ and $m_\rho$
as $f_\pi=92.4 {\rm MeV}$ and $m_\rho=776 {\rm MeV}$,
all the variables in the holographic QCD 
%
%like $\kappa$, $M_{\rm KK}$ and $e$ 
%
are uniquely determined as 
$\kappa\simeq 7.46\times 10^{-3}, M_{\rm KK}\simeq 948 {\rm MeV}, e\simeq 7.315.$
With these variables, we find reasonable mass $M_{\rm HH} \simeq 834{\rm MeV}$ 
and small 
%root-mean-square 
radius $\sqrt{\langle r^2\rangle} \simeq 0.37{\rm fm}$
for the hedgehog Brane-induced Skyrmion. 

To summarize, we have studied Brane-induced Skyrmions, {\it i.e.},
baryons in the holographic QCD 
with D4/D8/$\overline{\rm D8}$ multi D-brane system, 
and we have numerically obtained the hedgehog soliton solution 
of the holographic QCD.

The authors acknowledge the Yukawa Institute for Theoretical Physics at Kyoto University 
for useful discussions during the YKIS2006 on `New Frontiers in QCD'.
% were useful to complete this work. 

\end{document}